\documentclass[10pt,letterpaper]{article}
\usepackage[top=0.85in,left=1.25in,footskip=0.75in,marginparwidth=2in]{geometry}

\usepackage[utf8]{inputenc}

\usepackage{cite}

\usepackage{nameref,hyperref}

\usepackage{microtype}
\DisableLigatures[f]{encoding = *, family = * }

\raggedright
\usepackage{changepage}

\usepackage[aboveskip=1pt,labelfont=bf,labelsep=period,singlelinecheck=off]{caption}

\makeatletter
\renewcommand{\@biblabel}[1]{\quad#1.}
\makeatother

\usepackage{lastpage,fancyhdr,graphicx}
\usepackage{epstopdf}
\fancyheadoffset[L]{2.25in}
\fancyfootoffset[L]{2.25in}

\usepackage{color}
\usepackage{amsmath}
\usepackage{bm}
\definecolor{Gray}{gray}{.25}

\usepackage{graphicx}

\usepackage{sidecap}

\usepackage{wrapfig}
\usepackage[pscoord]{eso-pic}
\usepackage[fulladjust]{marginnote}
\reversemarginpar

\begin{document}
\vspace*{0.35in}

% title goes here:
\begin{flushleft}
{\Large
\textbf\newline{Online dynamic flat-field correction for MHz Microscopy data at European XFEL}
}
\newline
% authors go here:
\\
Sarlota Birnsteinova\textsuperscript{a,*},
Danilo E. Ferreira de Lima\textsuperscript{a},
Egor Sobolev\textsuperscript{a},
Henry J. Kirkwood\textsuperscript{a},
Valerio Bellucci\textsuperscript{a},
Richard J. Bean\textsuperscript{a},
Chan Kim\textsuperscript{a},
Jayanath C. P. Koliyadu\textsuperscript{a},
Tokushi Sato\textsuperscript{a},
Fabio Dall'Antonia\textsuperscript{a},
Eleni Myrto Asimakopoulou\textsuperscript{b},
Zisheng Yao\textsuperscript{b},
Khachiwan Buakor\textsuperscript{a,b},
Yuhe Zhang\textsuperscript{b},
Alke Meents\textsuperscript{c},
Henry N. Chapman\textsuperscript{c,d},
Adrian P. Mancuso\textsuperscript{a,e,f},
Pablo Villanueva-Perez\textsuperscript{b},
Patrik Vagovi\v{c}\textsuperscript{c,a}
\\
\bigskip
% \bf{a} European XFEL GmbH, Schenefeld Germany
a European XFEL GmbH, Schenefeld Germany
\\
b Synchrotron Radiation Research and NanoLund, Lund University Sweden
\\
c Center for Free-Electron Laser Science (CFEL), DESY, Hamburg Germany
\\
d University of Hamburg, Hamburg, Germany
\\
e Diamond Light Source Ltd., Harwell Science and Innovation Campus, Didcot, OX11 0DE United Kingdom
\\
f Department of Chemistry and Physics, La Trobe Institute for Molecular Science, La Trobe University, Melbourne, Victoria  Australia
\\
\bigskip
* sarlota.birnsteinova@xfel.eu

\end{flushleft}

\section*{Abstract}
The X-ray microscopy technique at the European X-ray free-electron laser (EuXFEL), operating at a MHz repetition rate, provides superior contrast and spatial-temporal resolution compared to typical microscopy techniques at other X-ray sources.
In both online visualization and offline data analysis for microscopy experiments, baseline normalization is essential for further processing steps such as phase retrieval and modal decomposition. In addition, access to normalized projections during data acquisition can play an important role in decision-making and improve the quality of the data. However, the stochastic nature of XFEL sources hinders the use of existing flat-flied normalization methods during MHz X-ray microscopy experiments.
Here, we present an online dynamic flat-field correction method based on principal component analysis of dynamically evolving flat-field images. The method is used for the normalization of individual X-ray projections and has been implemented as an online analysis tool at the Single Particles, Clusters, and Biomolecules and Serial Femtosecond Crystallography (SPB/SFX) instrument of EuXFEL. 

% now start line numbers
% \linenumbers

%----------------------
%
% The main body of the paper
%------------------------------------------------------
%       INTRO ........................................
%  ..................................................
% the * after section prevents numbering
\section{Introduction}

% intro to MHz microscopy
In the past few decades, X-ray microscopy was established as an important tool for the study of a large range of applications, from biology to material research, at third and fourth-generation X-ray synchrotron sources, where the achievable spatial resolution and sensitivity have been pushed to the limits~\cite{ChapmanBajt2015}. Recently, fast megahertz (MHz)-rate X-ray microscopy synchronized to individual pulses has been pioneered at the Advanced Photon Source (APS)~\cite{Fezzaa2008} and at the ESRF - the European Synchrotron~\cite{Olbinado2017, Olbinado2018}, which made it possible to image fast stochastic phenomena. MHz X-ray microscopy experiment with frame acquisition synchronized to individual X-ray pulses has been demonstrated recently at European XFEL (EuXFEL)~\cite{Vagovic19}. This development was enabled by the unique properties of X-ray Free-Electron Lasers (XFELs). In particular, EuXFEL is currently the only XFEL source providing intense pulses at MHz rates \cite{Decking2020}, which opens up the potential for novel methods and applications. EuXFEL provides two to four orders of magnitude more photons per pulse ($\sim 10^{12} - 10^{13}$) than the most brilliant synchrotron, i.e ESRF. EuXFEL also provides high spacial coherence, with a beam path of 1 km and a maximum of 4 $\mu$rad divergence, with photon energy up to 30 keV. The unique properties of EuXFEL come with unique issues, which complicate the acquisition of high-quality MHz microscopic data.  
XFEL's X-ray pulses are generated through the SASE process~\cite{milton01}, which leads to strongly stochastic spatial and temporal fluctuations.  
Another type of aberration appearing in microscopy data is often referred to as fixed-pattern noise. It originates from the different responses of the detector's pixels and imperfections in the optic systems, such as scintillator screens, lenses, etc. 
Ideally, image aberrations should  be corrected, so that only the signal variations originating from the X-rays' interaction with a sample are observed. Methods aiming to achieve this goal are often referred to as ``flat-field correction'' methods. 

Flat-field correction methods rely on using reference images averaged over many realizations of acquired images. Reference images acquired with beam illumination are referred to as ``flat-field images'', while images acquired without beam illumination are referred to as ``dark-field images''. However, as a stationary method, traditional flat-field correction, which assumes a flat-filed that is stable with time, is unable to sufficiently normalize stochastic effects, such as those stemming from SASE processes which cause variations in beam filed from XFEL pulse to pulse. %(?)   

There are several methods used for correcting both these effects~\cite{VanNieuwenhove:15,Hagemann21,Buakor:22,hodge22}.
The issue of normalization of a dynamically changing flat-field has been tackled by a method~\cite{VanNieuwenhove:15}, in which the reference flat-fields are represented in a latent lower-dimensional vector space and an ``effective'' flat-field is chosen per input image in the latent space. This method, called dynamic flat-field correction,  was introduced in Ref.~\cite{VanNieuwenhove:15} for synchrotron data. The application of a similar method was demonstrated on EuXFEL's full-field imaging data~\cite{Hagemann21} of high magnification and in combination with deep-learning approaches~\cite{Buakor:22}. 

In this paper, we present an \emph{online} method used at EuXFEL to perform dynamic flat-field normalization for MHz microscopy data. EuXFEL users often require fast feedback when performing an experiment, such that they may adapt their configurations to improve the quality of the data taken. In such conditions, obtaining corrected images during the experiment may be the deciding factor between a successful experiment, or a failure to obtain reliable data. The online flat-field correction method described in this paper is optimized to provide output to the user as data arrives, minimising the delay between data-acquisition and analysis. The method proposed here also provides an indication of whether the flat-field images collected as references are sufficient, or whether a new dataset must be collected due to changes in the experimental conditions. This helps to enhance the method's performance, as a dataset of flat-fields that sufficiently describes the illumination at a given moment of measurement has a crucial impact on the resulting quality of the normalization when affected by the fluctuations of SASE illumination. Moreover, the normalized data provided during experiments can be further re-used as input to other processing tools, for which pre-processed data are essential. 
Our main objective is to ensure high quality image data at high speeds. We are using a software tool provided by EuXFEL, enabling us to use online data and follow with the normalization procedure in near real-time.
The resulting online normalization tool for MHz microscopy was developed and implemented at the Single Particles, Clusters, and Biomolecules and Serial Femtosecond Crystallography (SPB/SFX) instrument of EuXFEL.
% how good it was? comparing to conven. method ?
Our online implementation based on the dynamic method led to a major improvement in the quality of corrected images compared to results using conventional flat-field correction approach. 

The manuscript is organised as follows: in Section~\ref{sec:normalisation}, we describe flat-field correction methods. The normalization algorithm and implementation are described in Section~\ref{sec:algorithm}. We discuss the results and the performance of implementation using test data in Section~\ref{sec:results}, followed by conclusion and discussion that are given in Section~\ref{sec:conclusion}.

%....................................................
%     METHODS      ..................................
%  ..................................................
\section{Flat-field correction \label{sec:normalisation}}
In this section, we briefly review of the conventional flat-field correction method.
We then expand on the dynamic flat-field correction method, which is implemented in this work. While we closely follow~\cite{VanNieuwenhove:15, Buakor:22}, differences in the methods are pointed out as needed.

A stationary, conventional flat-field correction removes the effects of spatially uneven illumination functions. Such effects may be caused by variations in the detector response in different pixels or modules, or effects arising from the scintillator and optics configurations. One assumes that there has been a previous data acquisition of $N_f$ flat-field images  $f_j$ and $N_d$ dark-fields $d_k$, where $j \in [1,N_f]$ and $k \in [1,N_d]$ refer to the reference images' index. In a stationary flat-field correction, one takes advantage only of the average flat-field and dark-field images, $\bar{f} = \frac{1}{N_f} \sum_j f_j$ and $\bar{d} = \frac{1}{N_d} \sum_k d_k$.

A simple procedure to correct the uneven detector response would be to correct each raw input image $p_i$ by calculating:
\begin{equation}
p_i^c = \frac{p_i - \bar{d}}{\bar{f} - \bar{d}}.
\label{eq:convFFCsample}
\end{equation}

As mentioned previously, due to the character of  SASE processes, noise is shot-to-shot-dependent, introducing a change in the intensity profile on the scintillator screen between shots. This behaviour cannot be captured solely by the average flat-field image, causing the conventional method to fail for such conditions.
% ..............................................
To correct for the stochastic effects observed in SASE sources such as EuXFEL, a dynamic flat-field correction substitutes the average flat-field with an effective flat-field $f_i^\prime$, which depends on the collected sample image $p_i$ itself. With this alternative procedure, the corrected image is

\begin{equation}
p_i^c = \frac{p_i - \bar{d}}{f^\prime_i - \bar{d}}.
\label{eq:dynamicFFCsample}
\end{equation}
In principle, the true illumination function may be any image. However, some simplifying assumptions allow for an approximate estimate of $f_i^\prime$. Firstly, it is natural to assume that such flat-field could be expressed as a function of the previously collected flat-fields $f_j$. While such a function may be extremely complex, a simple first-order approximation would be that $f_i^\prime$ is a linear combination of the collected flat fields. Namely, $f_i^\prime = \sum_j w_j f_j$. For the purposes of an online flat-field correction, this initial assumption leads to a fast and reliable implementation, which may be supplemented by further corrections offline.

While one may approach the issue of the approximation of an effective flat-field with as many weights as there are collected flat-fields, one may further simplify such a description by assuming that only a few collected flat-fields are the main contributors to the sum. A data-reduction technique could, therefore, be of use to reduce the number of coefficients required. Assuming the flat-fields collected are samples of a Gaussian distribution, one may use Principal Components Analysis (PCA)~\cite{Pearson1901,Hotelling36} to rewrite the linear flat-field combination as a linear combination of the principal components and keep only the components with the largest variance. Discarded components would then contribute little to the linear combination. In general, it is feasible to use $M$ components, where $M \ll N_f$.
In such a setting, we expand the effective flat-field $f^\prime_i$ as:
\begin{equation}
f^\prime_i = \bar{f} + \sum_m^M u_m w_{i\,m},
\label{eq:dynamicFF}
\end{equation}
where $u_m$ is the $m$-th principal component of the mean-subtracted flat-fields, $\{f_j - \bar{f}\}$, and $w_{i\,m}$ are free parameters to be identified online.

% % total variation minimization 
While a functional form is available to parameterize the effective flat-field $f_i^\prime$, one must determine a procedure to select the weights $w_{i\, m}$, such that the effective flat-field is uniquely chosen for a given sample image $p_i$. A constraint must be imposed on the corrected image for a meaningful choice of weights.

A key constraint is to preserve details originating from the interaction of X-rays with the sample. 
Such details appear through sharp transitions or edges in the image, separating the signal from the background content. The free parameters $w_{i\, m}$ must, therefore, preserve such edge effects, while removing spurious variations in the images caused by the uneven detector response. Additionally, one would like to remove spurious variations in the images added by an uneven flat-field. The regularization technique of Total Variation (TV) denoising is often used in image processing to achieve exactly those goals, by minimizing the TV of the corrected image, while aiming to obtain an image as similar as possible to the uncorrected one. 
Details concerning calculation of weights $w_i$ can be found in the following articles \cite{VanNieuwenhove:15,chambolle04}.

% summarize
The method described here consists of two separate steps. Initially, reference flat-fields and dark-fields are acquired and PCA is used to obtain the most relevant principal components of the flat-field dataset. During data-acquisition with a sample, for each individual frame, weights appropriate for the effective flat-field are found. % following Eq.~\ref{eq:w_i}. 
After such calculation, the final corrected frame is obtained by applying the weights in the linear combination in Eq.~\ref{eq:dynamicFFCsample}.

Such a procedure must be implemented in a way that it can be run online at adequately fast rates to keep up with the data-acquisition rate of EuXFEL. We expand on the technical implementation in the next section.

%.....................................................................  
%         Details & Implementation
%...................................................................
\section{Implementation \label{sec:algorithm}}
% % intro & how it all comes together
This section provides a schematic overview of the two steps of the main online normalization algorithm outlined in the previous section. Selected details of the implementation of the algorithm and specific features for visualization applied at EuXFEL are detailed here.

\begin{figure}%
    \centering
    {\includegraphics[width=9.5cm]{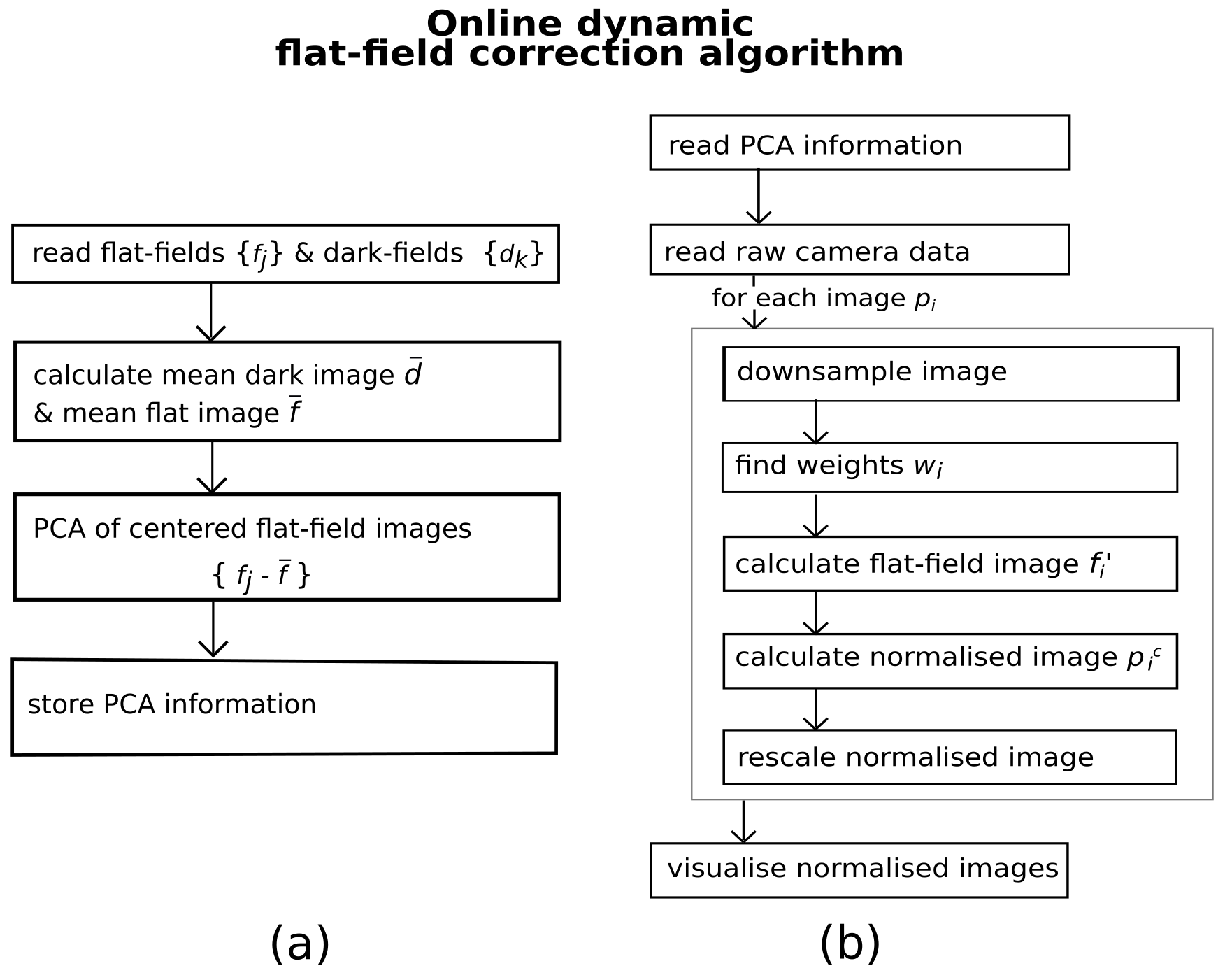} }%
    \caption{Algorithm overviews of the first part with principal component analysis of flat-field dataset (a) and the second part performing the online dynamic flat-field correction and visualization (b).}%
    \label{fig:schematics_both}%
\end{figure}

The schematic overview of the first part of the algorithm, employing PCA, can be found in Fig.~\ref{fig:schematics_both} (a), whereas Fig.~\ref{fig:schematics_both} (b) summarises the second part which estimates an individual flat-field for every sample image and visualizes the resulting normalized data.

Next, we list some of the specific functions and details used in the implementation of the algorithm. For the total variation minimization of the objective function a quasi-Newton algorithm is applied% , which is based on an approximation of the Hessian function using a limited memory BFGS matrix
~\cite{lbfgsb}. 
The precision parameter was tuned to maximise processing rates of the correction algorithm. Adjustments have been made to preserve the quality of the corrected images while increasing the processing speed. 

The speed of the algorithm is further increased by down-sampling all the images entering the minimization procedure \cite{VanNieuwenhove:15,Buakor:22}, causing also reduction of noise in corrected images. Here, in order not to introduce more time-consuming steps, we omitted any additional algorithms aiming exclusively for noise reduction, which may be included at later stages for offline processing.  

%Fig schema
%...........................................................
%    EuXFEL implementation
%...........................................................
% % idea & intro
This work demonstrates a novel implementation of the dynamic flat-field correction algorithm to be used at EuXFEL's infrastructure, which can be used as a normalization technique during experiments.
% % why? 
Moreover, considering the goal of user-friendliness of the online normalization tool, EuXFEL's software framework Karabo~\cite{heisen13} is used here for online visualization. 
One of the two implementations uses the \emph{metropc} framework and the \emph{extra-metro} package~\cite{fangohr18}, which is integrated within the Karabo framework and its Graphical User Interface (GUI). The \emph{metropc} framework allows for flexible adjusting of a script during an experiment. It allows the visualization of the corrected incoming data, in addition to raw sample images.  
The second implementation of the online dynamic flat-field correction algorithm is done using Karabo bridge~\cite{fangohr18} to access online data in the Karabo pipeline and perform parallelized analysis and further visualization using the Qt toolkit~\cite{pyqt_docu}. 
% \clearpage

%....................................................
%              RESULTS & EXPERIMENT   
%  ..................................................
\section{Results \label{sec:results}}
% % intro: the goal & proposed methods
We have described here the results obtained by our dynamic flat-field correction implementation at EuXFEL. We demonstrated our algorithm on a Venturi tube dataset~\cite{soyama16}. Data were acquired in March 2021 at the SPB/SFX instrument~\cite{Mancuso19} of EuXFEL with a photon energy of $9.3$ keV. A fast-frame-rate Shimadzu HPV-X2 camera, coupled to a $8 \mu m$ thick  LYSO:Ce scintillator, was used as an imaging system with a 10$\times$ magnification~\cite{Vagovic19}. Datasets consisting of 68 dark-field and 67 flat-field trains were taken to test our method. Each train consisted of 128 images with a size of $250\times 400$ pixels. The effective pixel size was $3.2 \mu m$. The exposure time of an image was ultra-fast and was given by the X-ray pulse duration translated into the latent image at the scintillator. The duration of the signal emitted by the scintillator was given by the scintillator emission time $\tau$ (LYSO:Ce $\tau \sim$ 40 ns). The frequency of acquired frames within a train was $1.128$ MHz and trains were collected at frequencies from $0.08-0.1$ Hz, limited by the camera idle time.
The performance of the dynamic flat-field correction algorithm  on the aforementioned dataset is demonstrated in this section.

% % PCA part ............................
The first step toward the normalization of online data is the processing of flat-field and dark-field data. Our aim was to acquire hundreds to thousands of flat-field images, which, at the current acquisition speed, requires up to $10$ minutes of measurements. 
After the first step of the algorithm, as described in Fig.~\ref{fig:schematics_both} (a), a preview of the cumulative sum of explained variance ratio of principal components can be viewed. This allows monitoring of the rate at which the selected number of principal components explain variance in the flat-field dataset. In addition, it enables the increase or decrease of the number of components before proceeding with the normalization of sample images.   

\begin{figure}% 
    \centering
    {{\includegraphics[width=7.5cm]{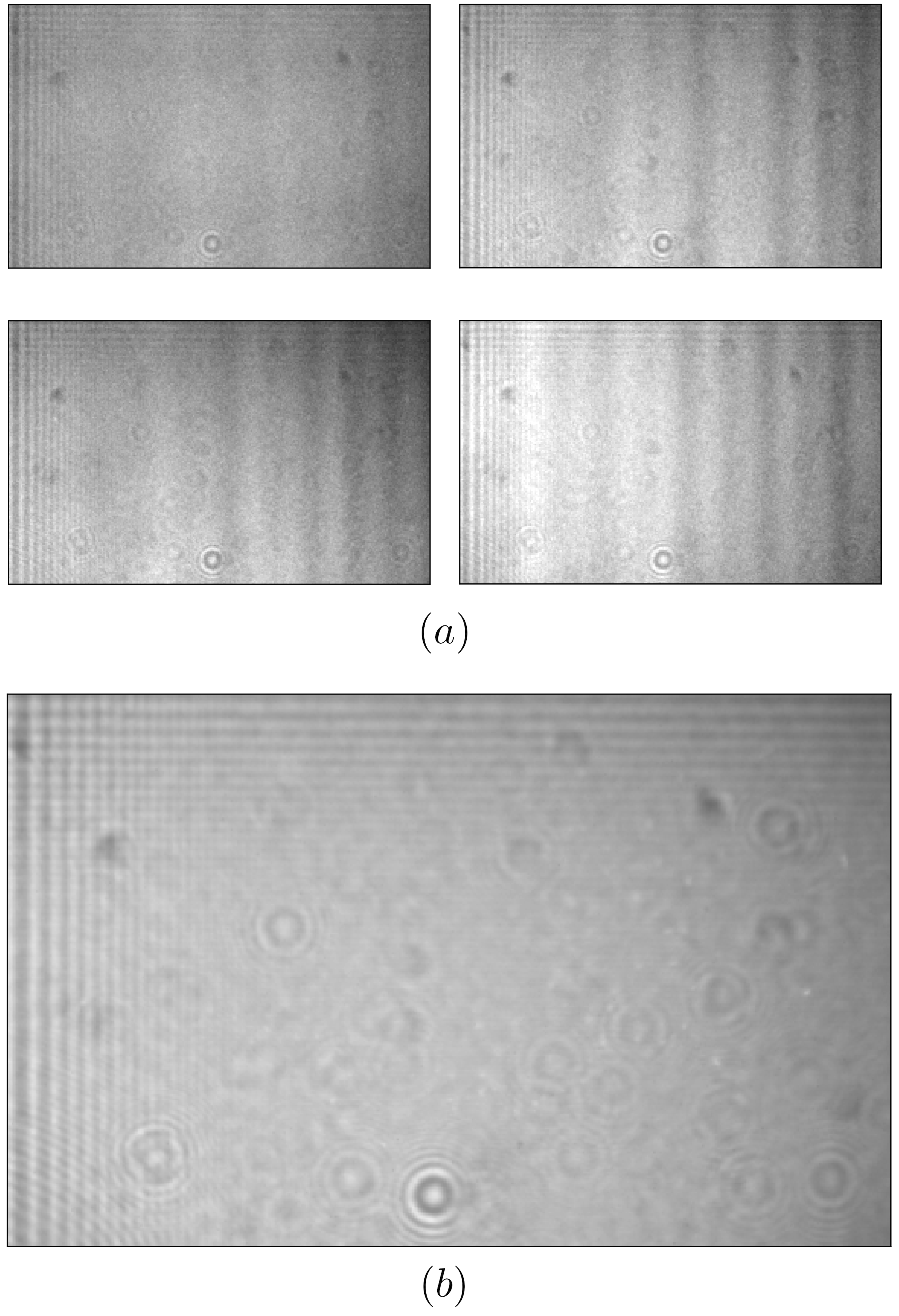} }}%
    \caption{Example instances~(a) of the test flat-field dataset consisting of approximately 10 000 images and the average flat-field image~(b) calculated over the whole test dataset.}
    \label{fig:exmpls_FF}%
\end{figure}

\begin{figure}%
    \centering
    {{\includegraphics[width=8.5cm]{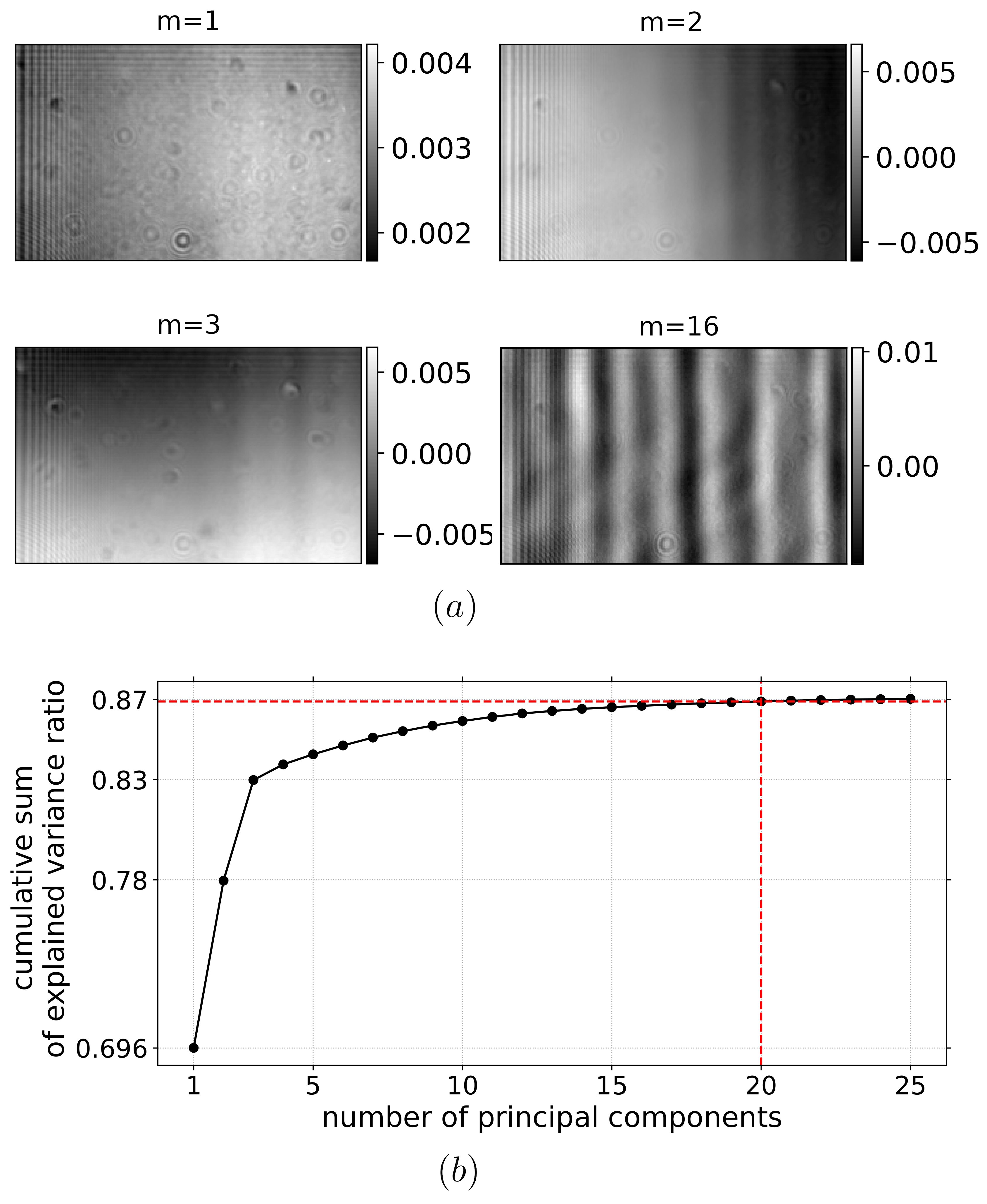} }}%
    \caption{An example of principal components $u_m$ for $m=\{1,2,3,16\}$ (a), obtained from the test flat-field data shown in Fig.~\ref{fig:exmpls_FF}. The cumulative sum of explained variance ratio of the first 25 principal components is depicted in subfigure (b).}%
    \label{fig:pca_modes_example}%
\end{figure}

% PCA on example data
The flat-field dataset consists of approximately 10 000 flat-field images with a few instances shown in Fig.~\ref{fig:exmpls_FF} (a), part~(b) shows the mean flat-field image calculated over the whole dataset. In Fig.~\ref{fig:pca_modes_example}, selected principal components are depicted for the index of principal component $m=\{1,2,3,16\}$. The first principal component, labeled $m=1$, is similar to the mean flat-field image in Fig.~\ref{fig:exmpls_FF} (b) and explains the largest fraction of the variance in the dataset. The following two components, $m\in \{2, 3\}$, capture the intensity variation in horizontal and vertical directions. %\in [4--18]
Components $m \leq 4$, show mostly horizontal changes in intensity with pronounced stripe features, as depicted on the component $m=16$. 
The cumulative sum of explained variance ratio of the flat-field dataset is shown in Fig.~\ref{fig:pca_modes_example} for the first $25$ principal components.
% Figs flat & PCA
Figure~\ref{fig:pca_modes_example} (b) shows that the first three components explain approximately $83\%$ of the variance. Adding more components does not increase it significantly, and after component $3$ the increases in cumulative sum of explained variance are minor. However, as will be discussed in more detail later in this chapter, in order to correct for the majority of vertical features in sample images, one needs to include also those components with seemingly low values of explained variance, as they obtain variations needed for a high-quality correction. 

\begin{figure}%
    \centering
    {{\includegraphics[width=8.5cm]{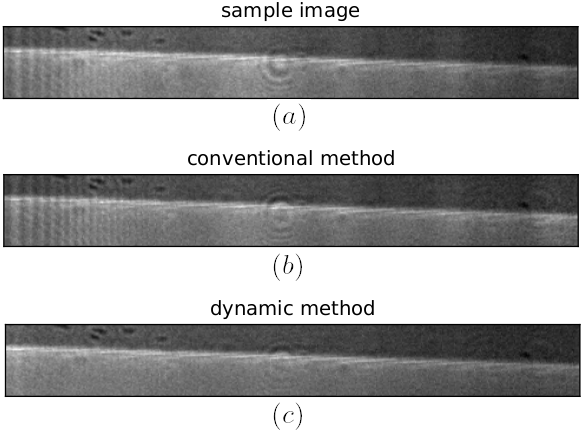} }}%
    \caption{Comparison of a sample image (a) without any normalization, conventionally flat-field corrected image (b) and dynamic flat-field corrected data (c). Only the bottom section of images ($(50,400)$ pixels) is shown here from originally sized images ($(250,400)$ pixels).}%
    \label{fig:dFFc_compar_all}%
\end{figure}

% % dFFc part ..........................................
After identifying the most important components describing flat-field illumination, one can continue with the normalization procedure of sample images described in Section~\ref{sec:algorithm}. To better visualize the performance of dynamic flat-field correction on the test data, in Fig.~\ref{fig:dFFc_compar_all} we plot the sample image (a), its counterparts corrected by both conventional (b) and dynamic method (c). The default value for a number of principal components taken into consideration here is 20. 
% Fig raw vs. corrected 
Fig.~\ref{fig:dFFc_compar_all} illustrates that the dynamic method~(c) more successfully removes vertical features than the conventional correction method~(b).
We can notice vertical features along the whole width of an image changing position from frame to frame, which is captured in examples~(a) Fig.\ref{fig:exmpls_FF}. Moreover, the lack of those features on the average flat-field image~(b) Fig.\ref{fig:exmpls_FF} causes the inability of conventional method successfully remove such artefacts from acquired images. Even fine stripes, appearing on average flat-field~(b) Fig.\ref{fig:exmpls_FF} on the left side, are not removed using conventional normalisation method, in Fig.\ref{fig:dFFc_compar_all} (b), due to the slight change of their position during measurement. Another challenging detail to eliminate is the circular spot from the middle of the sample image~(a). Similar to the fine stripes from the previous example, better, however not perfect, reduction of the spot is reached only with the dynamic method~(c) Fig.\ref{fig:exmpls_FF}.

% % quantitative comparison
The quality of flat-field corrected images was mostly assessed visually. To estimate the method's performance quantitatively, we calculated 
pixel value spread (pvs) as a deviation of an image from its average pixel value
\begin{equation}
 \text{pvs} = \sqrt{\frac{\sum_{l}^{N^{\prime}} (p_l - \mu)^2 }{N^{\prime}}},\,\, \mu=\sum_{l}^{N^{\prime}} p_l
\end{equation}
where index $l$ iterates over pixels of image $p$. We were using only ''flat'' parts of images, without any sample, in order to estimate the amount of noise present in an image. Higher pvs values are expected for uncorrected images with uneven intensity profile and lower values for images, which were corrected in a way, that reduces most of the uneven structures and noisy pixels. 
 
\begin{figure}% 
    \centering
    \includegraphics[width=8.5cm]{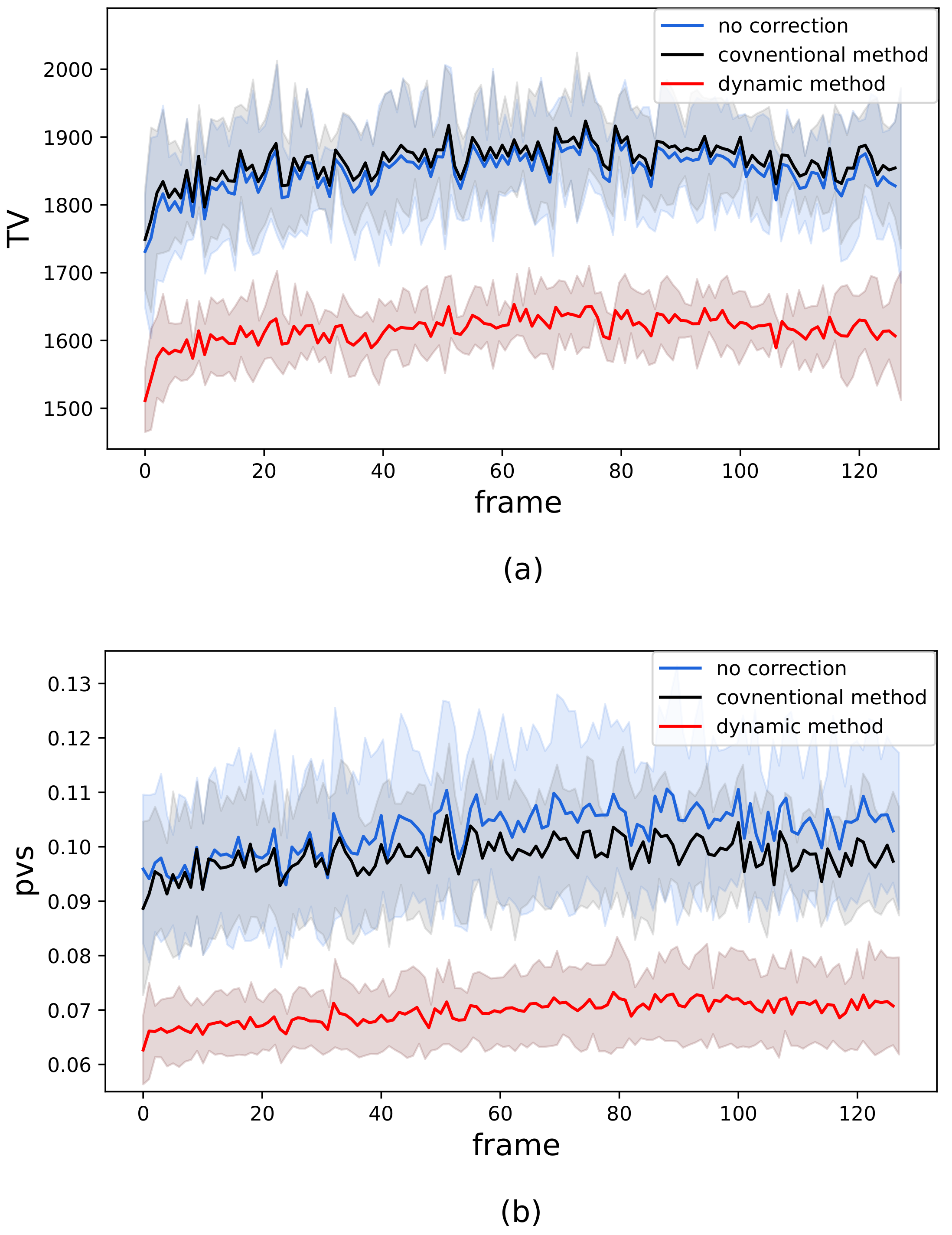}
    \caption{Comparison of total variation~(a) and pixel value spread~(b) for uncorrected images and images corrected  by conventional and dynamical method. Values of TV and pvs are calculated depending on their position within a train, where the maximum frame number is 128. Both TV and pvs for each frame number are averaged over 15 trains and train-to-train changes are captured by their standard deviation expressed as a colored area.}%
    \label{fig:comparison_tv_rmse_bothMethods}%
\end{figure}

In order to assess the quality of correction of images of the same position within a train we calculated TV and pvs for each of 128 images acquired in one train and averaged over 15 trains. Average values for both variables and their standard deviations are shown in Fig.\ref{fig:comparison_tv_rmse_bothMethods}. While the pvs and TV value varies for each frame, one may see in Fig.~\ref{fig:comparison_tv_rmse_bothMethods} that the dynamic flat-field corrected images have a consistently lower TV (a) and spread (b), suggesting the lowering of noise level in the chosen sample-free part of the images. In contrast, the conventional flat-field correction shows very little benefit, in comparison to no correction at all, as for TV and pvs values no statistically significant difference was found for uncorrected image and images corrected using conventional method. This result was expected based on the visual comparison in Fig.\ref{fig:dFFc_compar_all}, in which no visible removal of vertical features was observed. Standard deviation expressed as colored area for uncorrected images and both normalization methods in Fig.~\ref{fig:comparison_tv_rmse_bothMethods} describes train-to-train changes in TV and pvs value for a given frame in dataset. On average, the pvs value for a dynamically corrected image was found to be 0.069, while the spread for conventionally corrected images was 0.098. Uncorrected images have a pvs of 0.102.

\begin{figure}% 
    \centering
    \includegraphics[width=8.5cm]{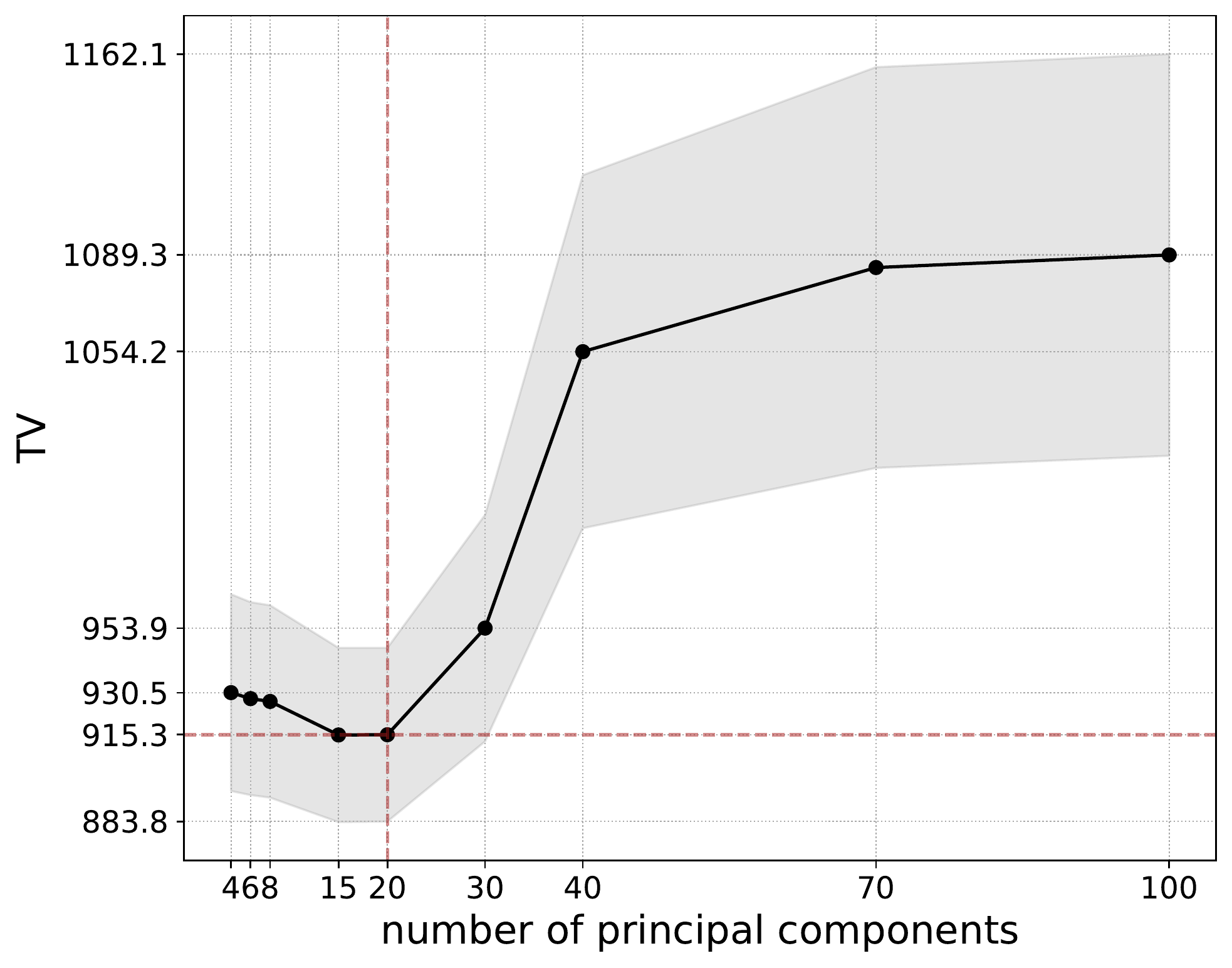}
    \caption{Dependence of TV of images corrected by the dynamical method using a different number of principal components. Values are calculated for one train and are averaged over 128 images with gray colored area given by standard deviation of frame-to-frame changes within a train.}%
    \label{fig:comparison_TV_parameters}%
\end{figure}

\begin{figure}% 
    \centering
    \includegraphics[width=8.5cm]{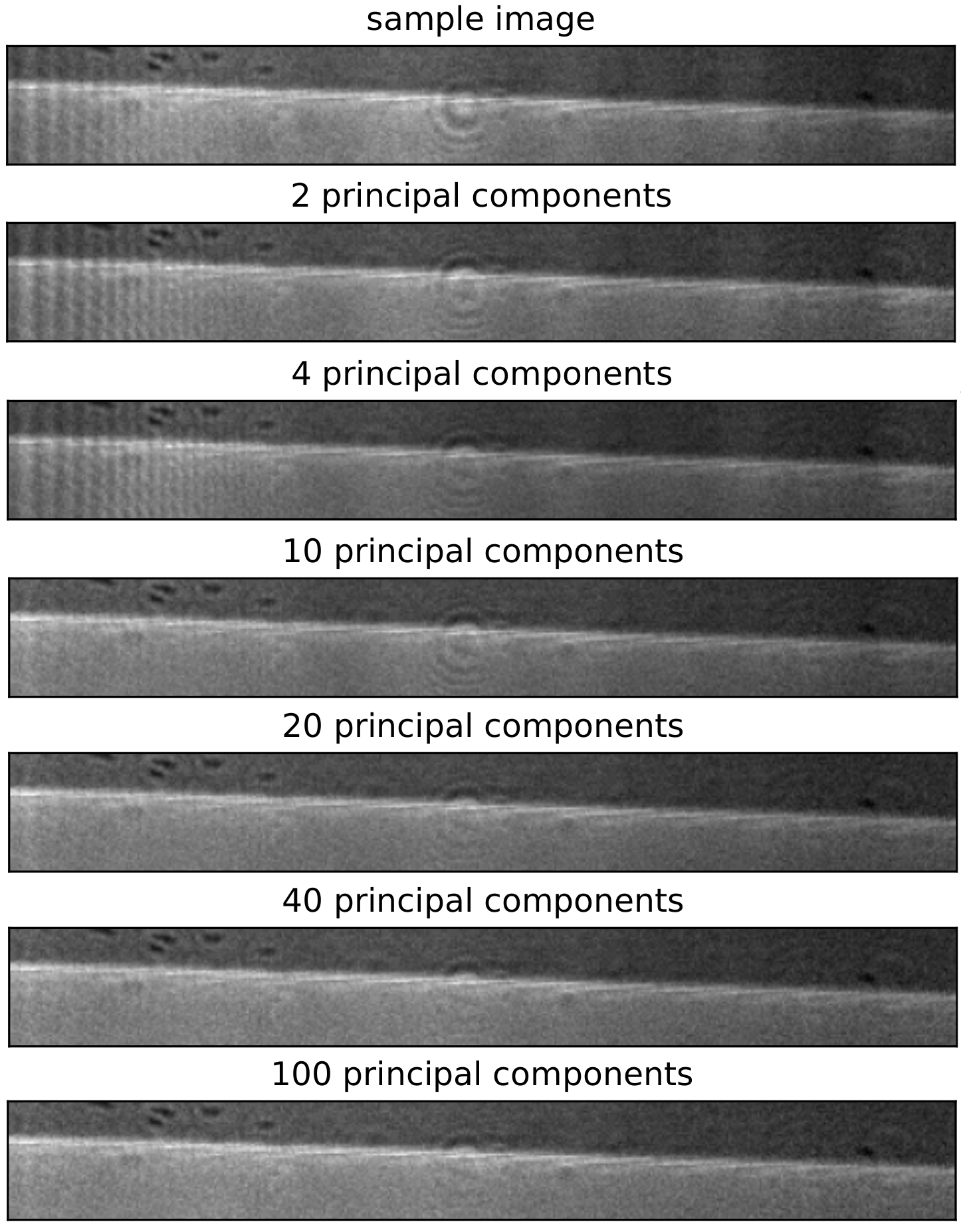}
    \caption{Comparison of uncorrected sample image and flat-field corrected images using the dynamical method with $[2,4,10,20,40,100]$ principal components.}%
    \label{fig:comparison_dffc_no_components}%
\end{figure}

% number of principal component
Additionally, we studied the quality of normalized images estimated by TV values as a function of a varying number of principal components. Resulting corrected images were also assessed visually.
Fig.~\ref{fig:comparison_TV_parameters} describes changes in TV value due to a varying number of principal components used in the normalization procedure. One can notice a decrease in average value of TV for numbers of principal components $15$ and $20$, followed by an increase starting after the $20$ components mark.
For a visual comparison, Fig.~\ref{fig:comparison_dffc_no_components} provides examples of flat-field corrected images using a varying number of principal components in the normalization algorithm.  
Fig.~\ref{fig:comparison_dffc_no_components} illustrates that including 2-10 principal components in the normalization procedure leads to non-optimal flat-field correction results. 
Fine line structures are still visible on the left part of images for 2 and 4 component cases. Vertical features along the whole width are not fully eliminated using 10 principal components.
Increasing number of principal components up to 10 results in decreased visibility of vertical features, which also supports observation of decreasing TV values in Fig.~\ref{fig:comparison_TV_parameters}. The circular spot in the middle is becoming less noticeable with higher number of principal components, although can be detected on all corrected images up to the case using 100 components.
Including around twenty principal component seems to remove majority of the unwanted features and results for 40 and 100 components in Fig.~\ref{fig:comparison_TV_parameters} are not leading to visibly better results. TV values in Fig.~\ref{fig:comparison_TV_parameters} are growing after 20 components, which may be caused by overfitting. %? 
The results listed in Figs.~\ref{fig:comparison_TV_parameters} and~\ref{fig:comparison_dffc_no_components} indicate that the optimal number of principal components is around $20$, where most of the features appearing on flat fields are removed.  

% time-wise conclusion
Considering the idle time of the camera, our goal was to reach a correction before the next acquired train arrives. The idle time of the camera is around $15$s, meaning the camera is able to acquire one in 150 of EuXFEL's trains. This consideration does not take into account set up time spent for the analysis of flat- and dark-field datasets, as this process usually takes a few minutes and needs to be repeated on an hourly basis, whenever new flat- and dark-fields are taken. The proposed algorithm and both its implementations mentioned in Sec.~\ref{sec:algorithm} were able to be faster and keep up with the volume of incoming data. A parallelized version reached correction time of approximately $1$s for a dataset consisting of one train with $128$ images. Tests were performed at online cluster INTEL, Gold-6140 CPU @ 2.30 GHz, 768G. Moreover, the processing speeds should be able to sustain the load with anticipated updates, which aim to shorten the idle time of the camera. The current performance of the algorithm is shown to be suitable for its use during experiments as a near real-time analysis tool.


% RMSE vs. pc and factr vs. time 
% comparinson dffc & # componnets

\clearpage

\section{Conclusion and Discussion \label{sec:conclusion}}

% % intro again 
We have demonstrated a method for dynamic flat-field correction on MHz repetition rate EuXFEL data. It has been shown that, for X-ray MHz microscopy data~\cite{Buakor:22}, the dynamic method results in an improved correction compared to the conventional flat-field correction method, taking into account temporal variations in the flat-field dataset, which is characteristic for XFEL facilities.
To compare the performance of conventional and dynamic methods applied to test data, we have calculated total variation, and the pixel value spread for the area of images without any sample. Both quantitative comparison and visual assessment have shown improved results with the dynamic normalization method compared to conventional method. The normalisation algorithm has been implemented online to visualize data and flat-field corrected data during experiments at the SPB/SFX instrument of European XFEL. 
The implementation was able to correct and visualize incoming data in near real-time, before the next data from the MHz-frame-rate camera arrived. 
% what does it mean for future experiments
In future, we plan to extend the use by implementing new data analysis tools such as data decomposition methods and adding more features to the GUI. 
%\clearpage

% \section*{Supporting Information}
% If you intend to keep supporting files separately you can do so and just provide figure captions here. Optionally make clicky links to the online file using \verb!\href{url}{description}!.

% %These commands reset the figure counter and add "S" to the figure caption (e.g. "Figure S1"). This is in case you want to add actual figures and not just captions.
% \setcounter{figure}{0}
% \renewcommand{\thefigure}{S\arabic{figure}}

% % You can use the \nameref{label} command to cite supporting items in the text.
% \subsection*{S1 Figure}
% \label{example_label}
% {\bf Caption of Figure S1.} \textbf{A}, If you want to reference supporting figures in the text, use the \verb!\nameref{}!. command. This will reference the section's heading: \nameref{example_label}.

% \subsection*{S2 Video}
% \label{example_video}
% {\bf Example Video.} Use \href{www.youtube.com}{clicky links} to the online sources of the files.

%\clearpage

\section*{Acknowledgments}
This work was performed within R\&D project “MHz microscopy at EuXFEL: From demonstration to method”, 2020 - 2022.

% \nolinenumbers

%This is where your bibliography is generated. Make sure that your .bib file is actually called library.bib
\bibliography{library}

%This defines the bibliographies style. Search online for a list of available styles.
% \bibliographystyle{abbrv}
\bibliographystyle{ieeetr}

\end{document}